\documentclass{PoS}
\usepackage{rotating}

\title{Cherenkov Telescope Array Science: Multi-wavelength and multi-messenger perspectives}

\ShortTitle{CTA MWL \& MM Perspectives}

\author{\speaker{Ulisses Barres de Almeida}\\
        Centro Brasileiro de Pesquisas F\'{i}sicas, Rua Dr. Xavier Sigaud 150, 22290-180 URCA, Rio de Janeiro (RJ), Brazil\\
        E-mail: \email{ulisses@cbpf.br}}

\author{representing the CTA Consortium\footnote{see http://www.cta-observatory.org/consortium\_authors/authors\_2019\_07.html for full author list}}

\abstract{The Cherenkov Telescope Array (CTA) will be the major global observatory for VHE gamma-ray astronomy over the next decade and beyond. It will be an explorer of the extreme universe, with a broad scientific potential: from understanding the role of relativistic cosmic particles, to the search for dark matter. Covering photon energies from 20 GeV to 300 TeV, and with an angular resolution unique in the field, of about 1 arc min, CTA will improve on all aspects of the performance with respect to current instruments, surveying the high energy sky hundreds of times faster than previous TeV telescopes, and with a much deeper view. The very large collection area of CTA makes it an important probe of transient phenomena. The first CTA telescope has just been inaugurated in the Canary Islands, Spain, and as more telescopes are added in the coming years, scientific operation will start. It is evident that CTA will have important synergies with many of the new generation astronomical and astroparticle observatories. In this talk we will review the CTA science case from the point of view of its synergies with other instruments and facilities, highlighting the CTA needs in terms of external data, as well as the opportunities and strategies for cooperation to achieve the basic CTA science goals.}

\FullConference{36th International Cosmic Ray Conference -ICRC2019-\\
		July 24th - August 1st, 2019\\
		Madison, WI, U.S.A.}

\begin{document}

\section{Introduction}

The Cherenkov Telescope Array (CTA) will be an observatory with two arrays of imaging atmospheric Cherenkov telescopes (IACTs), one in each hemisphere, for complete sky coverage. Composed of a large number of telescopes and three different classes of telescopes -- Large, Medium, and Small-Sized Telescopes (LSTs, MSTs, SSTs) --  CTA will operate in a wide energy range, starting at around 20 GeV, to beyond 300 TeV, and will improve the sensitivity level of current instruments by an order of magnitude at 1 TeV.  These unique characteristics of sensitivity and energy coverage give CTA access to unprecedented science in the very high-energy (VHE) domain. The CTA telescopes will have Cherenkov cameras with wide fields of view, between $>4.5^{\circ}$ and  $>8^{\circ}$ depending on telescope type.  Allied with an improved angular resolution, this will allow for the excellent imaging of extended sources, and performance of Galactic and extragalactic surveys. The large number of telescopes gives array operation flexibility, with the possibility for simultaneous observations of multiple fields.

In the time domain, CTA will be a uniquely powerful instrument for the exploration of the variable universe. The energy range of CTA science is dominated by extreme variability, and CTA is prepared to respond to external alerts with great efficiency, thanks to a maximum repointing time of 30 to 90 s, depending on the telescope open under consideration. The wide field of view and the unprecedented sensitivity of CTA (specially on short integration timescales) will likely make it into an efficient VHE transient factory, providing significant detections of serendipitous transient sources or variable emission episodes. A planned real-time analysis pipeline which will monitor the CTA field of view will also allow it to quickly send alerts for external observatories or to trigger deeper internal follow-ups of interesting and rare events, such as gamma-ray bursts (GRBs).

CTA will function as an open observatory. Nevertheless, during the first decade of its operation, part of the CTA observation time (about 40\%) will be devoted to a Core Programme, consisting of a number of major legacy projects which will be conducted by the CTA Consortium with guaranteed time~\cite{CTA2019}. The Core Programme is comprised of a series of Key Science Projects (KSPs) of particular scientific value and which demand considerable observation time for their completion. The efficient achievement of the broad-ranging science questions addressed by the CTA KSPs depends strongly on external data, which will complement the information provided by CTA observations. This contribution is concerned with giving a general overview of such demands, aiming at developing future strategies to maximise data access and the global scientific return of CTA's multi-wavelength (MWL) and multi-messenger (MM) synergies. 

As a matter of fact, CTA science will have important interfaces with many of the new generation of astronomical and astro-particle observatories. Along with CTA, new major facilities are coming into operation in this or the next decade, throughout the electromagnetic spectrum -- e.g., SKA in radio waves, the giant optical telescopes E-ELT, TMT and GMT, amongst others -- which will revolutionise their respective fields. New multi-messenger facilities, or planned upgrades of current ones, will take the recent, early achievements of these new fields of observational astronomy a step forward. In a two-way relationship, CTA science will majorly benefit from, and will greatly influence, key results and discoveries by these new telescopes. 

In the following, we will describe the CTA Science Programme and highlight its MWL and MM synergies and requirements, with some focus on transient sources towards the end.

\section{The Galaxy with CTA}

\begin{figure}[!htb]
        \center{\includegraphics[width=\textwidth]
        {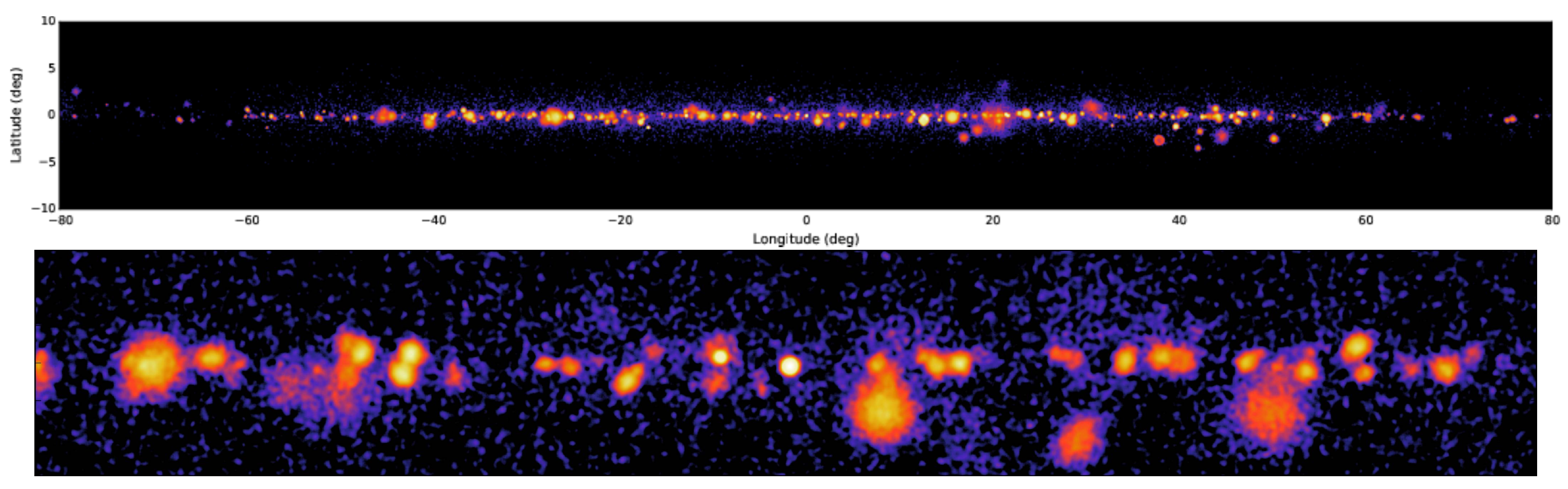}}
        \caption{\label{fig:gps} (Top) Simulated CTA image of the Galactic plane inner region, $-80^{\circ} < l < 80^{\circ}$, adopting the proposed GPS KSP observation strategy and a source model incorporating both SNR and PWNe populations, as well as diffuse emission. (Bottom) Zoomed image of a sample $20^{\circ}$ region in Galactic longitude~\cite{CTA2019}.}
      \end{figure}

The foundation for Galactic science with CTA will be provided by the {\bf Galactic Plane Survey} (GPS), which will fulfil a number of goals, of which the most basic one will be to provide a census of Galactic very high-energy (VHE) gamma-ray source populations, such as supernova remnants (SNRs), pulsar-wind nebulae (PWNe), and binary systems, substantially increasing the source count thanks to the CTA improved sensitivity (see Figure~\ref{fig:gps}). In performing the GPS-- which will consist of a deep survey down to $\sim 2$ mCrab of the inner Galaxy and the Cygnus regions, plus a shallower survey (down to $\sim 4$ mCrab) of the entire Galactic plane -- determining the properties of the Galactic diffuse emission will also be an important goal. Finally, the survey will produce a multi-purpose, legacy data set of long-lasting value to the entire astronomical community.

Special attention will be dispensed to the inner Galaxy within the GPS. Deeper exposures of the inner few degrees of the Galaxy will be performed for a detailed study of the Galactic Centre. These observations will be complemented by an extended survey to explore regions not yet covered by existing VHE telescopes, at high latitudes, to the edge of the bulge emission, including the base of the Fermi Bubbles and interesting sources such as the Kepler SNR. The core goal of the {\bf Galactic Centre Survey} is to provide unprecedented spatial and spectral sensitivity of this crucial regions of the Galaxy which may allow us to identify the central source~\cite{Aharonian2009} and decide between the models proposed to explain the extended emission~\cite{Aharonian2006}, thus providing a deeper understanding of the capability for cosmic ray acceleration in the Galaxy~\cite{HESS2016}. The Galactic Center is also a top Dark Matter target for CTA.

The search for the sources of petaelectronvolt protons (the so-called {\bf PeVatrons}) will be another main focus of the GPS. It is well known that SNRs are able to satisfy the cosmic-ray energy requirement~\cite{Hillas2005}, via diffusive shock acceleration at the expanding SNR shocks~\cite{Drury1983}. However, it is still unclear whether or not they can act as cosmic-ray PeVatrons, given uncertainties in the mechanisms of magnetic field amplification~\cite{Bell2004}. The observation of {\bf Star-Forming Regions} with CTA will help unveil the relation between cosmic rays and star formation.  Among the prime regions for such study are the Carina and Cygnus regions and the massive stellar cluster Westerlund 1. Outside the Galaxy, the Large Magellanic Cloud  (LMC) will be target of a dedicated survey.

\section{The AGN and Extragalactic Science Case}

\begin{figure}[!htb]
        \center{\includegraphics[width=\textwidth]
        {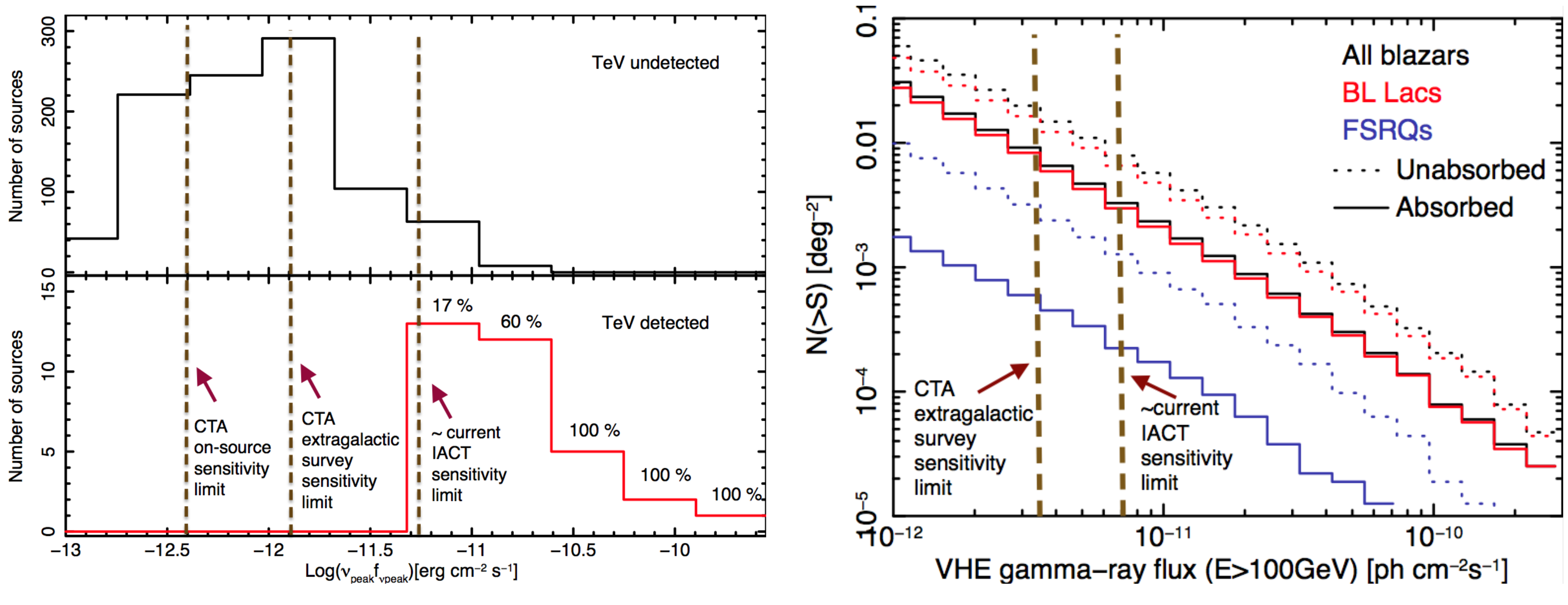}}
        \caption{\label{fig:xgal} Predictions for the number of blazars on the sky in the GeV-TeV domain. (Left) Source counts versus peak synchrotron energy flux. (Right) LogN-LogS distribution showing the expected source counts from blazars as a function of the integral gamma-ray flux above 100 GeV. Figure taken from~\cite{CTA2019}.}
      \end{figure}

From the side of the extragalactic science, CTA will perform the first blind {\bf Extragalactic Survey} at VHEs, covering 25\% of the sky. Like the GPS, the primary goal of the survey is to construct an unbiased VHE extragalactic source catalogue, with an integral sensitivity limit of about $\sim$ 6 mCrab above 125 GeV. At the moment, there are about 60 extragalactic sources seen with current IACTs, most of which are BL Lacs. This sample is, however, strongly biased, since most observations were made in flaring states, based on triggers from lower frequencies.

{\bf Active Galactic Nuclei} (AGN) represent a unique tool to probe the physics of extreme environments. They are also promising objects in which to look for signatures of ultra high-energy cosmic rays (UHECRs), thanks to the recent multi-messenger connection to IceCube TeV neutrino events~\cite{IceCube2018}. They are also bright beacons serving as probes of cosmological evolution through constraints on photon and magnetic fields along the line of sight. Apart from an unbiased view of their transient character, the studies of AGN performed with the CTA extragalactic survey will allow us, for the first time, to get an unbiased glimpse of the highest energy population of these objects, and probe new source classes, such as the putative "extreme blazars", with spectral energy distribution (SED) peaks above 1 TeV~\cite{Arsioli2018}. Thanks to its increased sensitivity and lower energy threshold, CTA has also the potential to substantially improve the redshift coverage and increase the population of known VHE sources at high redshifts (Figure~\ref{fig:xgal}). The highest redshift of the current sample is $\sim$ 0.9, but extrapolations from the population of Fermi-LAT objects show that over 200 additional blazars, as well as several radio galaxies, should be visible by CTA, up to $z \sim$ 2~\cite{Sol2013} .

Finally, {\bf Galaxy Clusters}, constitute another primary extragalactic science target for CTA. Their importance and interest relies on the fact that they are expected to be reservoirs of cosmic rays. The detection of diffuse synchrotron radio emission in several clusters confirms the presence of cosmic-ray electrons and magnetic fields permeating the intra-cluster medium (ICM). And while there is no direct proof for proton acceleration yet, gamma rays can probe it, as such particles are expected to give out high-energy gamma-ray emission through neutral pion decay. 

\section{MWL synergies for CTA}

In studying cosmic VHE non-thermal emission, CTA will probe the cosmic ray particle population and the extreme environments which produce them. By nature, such non-thermal emission spans many orders of magnitude in energy, covering the broad band electromagnetic spectrum. To understand the particle populations responsible for the observed emission, it is necessary therefore to have a combined view of the entire SED of the sources. Such a coordinated view is also necessary for the broader study of the various source properties and source populations detected by CTA. In particular, MWL correlations, and the cross-matching between MWL catalogues, are the means to enable source identification and to provide a detailed review of their environment, especially in the Galaxy, where fields are considerably more crowded, and source-ISM interactions are very important. The detailed study of source variability not only demands the combination of large MWL data sets, but also the temporal coordination, or simultaneous observations, between many facilities, with a significant time demand on CTA for follow-up observations of external alerts. Finally, in many cases, source distance estimations, both in the Galaxy and beyond, are only possible in a context where MWL data is available. Figure 3 provides an overview of the MWL / MM synergies.

Immediate MWL synergies for CTA science come from {\bf X-rays}, which have origin in extreme environments which can be readily associated with particle acceleration sites, such as shocked matter, accretion, or high-velocity outflows. The study of X-ray thermal emission provides otherwise unaccessible plasma properties and energetics. As a complement to the VHE observations, X-ray data are important for providing high-resolution images of extreme environments, with greater sensitivity than achievable by CTA. Key facilities are the spectro-imaging missions Chandra and XMM-Newton, and the timing observatories Suzaku and Swift, and NuStar, in the hard X-ray domain. Such telescopes are expected to continue functioning during the first few years of CTA operations, but new missions will guarantee continued synergies with CTA such as eROSITA, which  will provide an all-sky imaging survey in the 2-10 keV band. 

Thanks to excellent angular resolutions, {\bf radio and sub-millimetre} observations offer the best support in localising the particle acceleration zones. Radio measurements can also provide unique constraints on the magnetic fields at the sources through polarimetric and Faraday rotation measures, and provide pulsar ephemerides which allow for the search of pulsed emission in the VHE domain. In terms of facilities, radio astronomy is entering a phase of rapid development. At the low-frequency range, below 140 MHz, LOFAR is the main example. Of special notice is the future Square Kilometre Array (SKA), whose first phase of operations  will coincide with CTA's science verification phase, with full operations expected from 2024. SKA will have unprecedented sensitivity and angular resolution, with very large field of view capabilities enabled by phased-array technology. SKA pathfinders, such as the Australian ASKAP and UTMOST (100s MHz), and MeerKAT (GHz) in South Africa, are already in operation, and present important synergy potential with CTA. In the mm and sub-mm bands it is worth mentioning facilities such as Mopra (Australia), APEX and Nanten2 (Chile), as well as the already-operational Atacama Large Millimetre Array (ALMA), for supporting the study of diffuse emission around accelerator sites.

\begin{sidewaysfigure}[!htbp]
     \includegraphics[width=\textwidth]{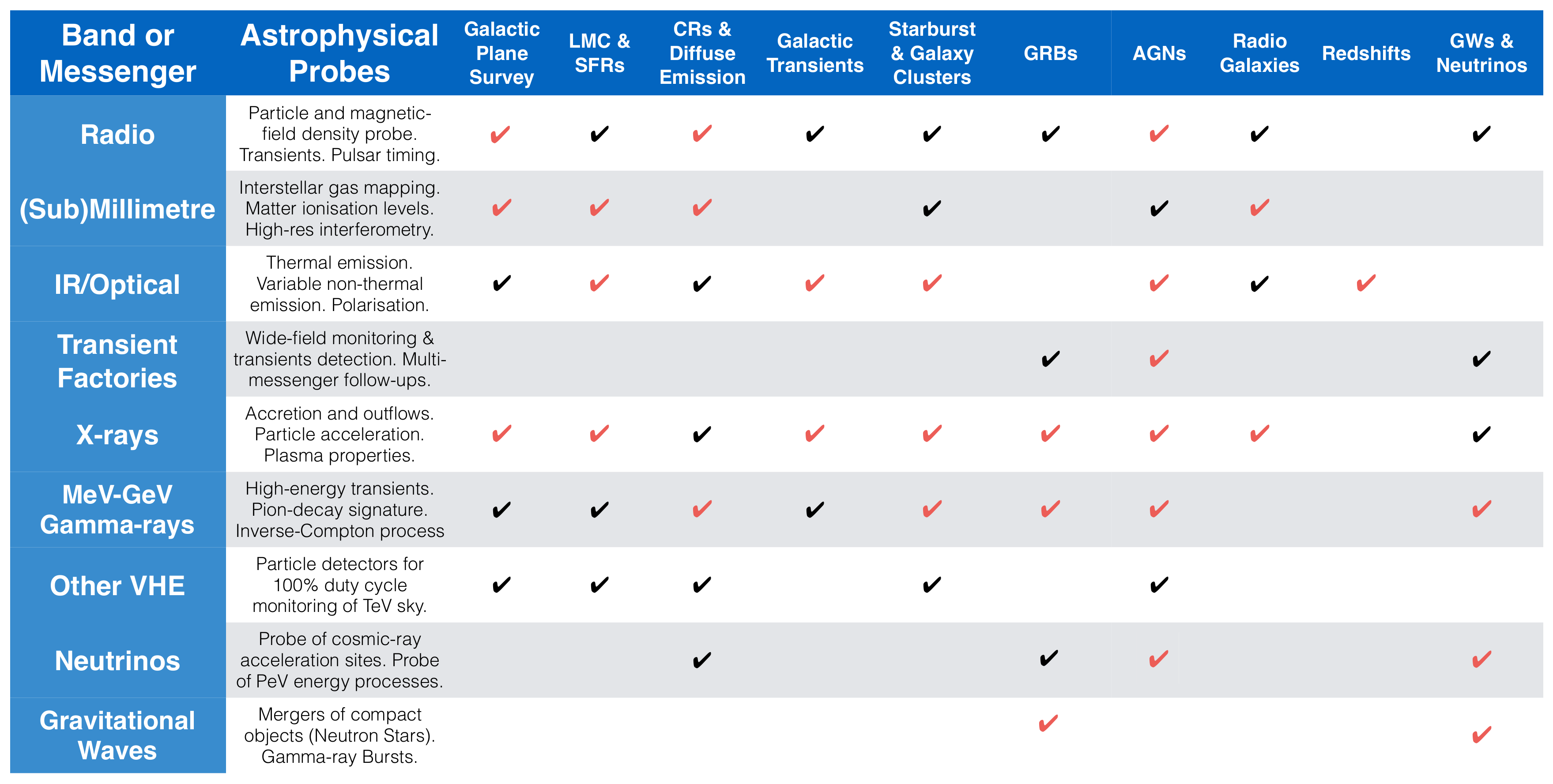}
\caption{\label{fig:table} Matrix of CTA Science Cases and associated MWL / multi-messenger synergies. The science cases listed refer to the core science programme of CTA, to be developed within the Consortium proprietary time. Some comments on the astrophysical capabilities from each band are also added. Ticks marked in red are to indicate the principal synergies of each science case.}
\end{sidewaysfigure}



Intensely variable, non-thermal synchrotron emission in {\bf optical/IR} is observed from a wide class of transient sources, such as blazars, microquasars and pulsar-wind nebulae, all of which are gamma-ray emitters. The advantage of the optical domain as a counterpart to CTA science is the fact that the large number of small-to-medium facilities worldwide provide a secure network of telescopes to complement CTA science in a very flexible way. The relatively low cost of such facilities also opens the way for potential Optical Support Telescopes (OSTs) on site. The Consortium is currently studying the feasibility of installing two 1-m class telescopes with polarimetric capability on each site to dedicatedly service CTA science, specially for transient source monitoring and coordinated multi-band campaigns, which demand a lot of coverage time from external facilities. In terms of telescopes, the optical domain will see a wealth of new, very-large ground-based facilities coming online in the next decade, such as the European Extremely Large Telescope and the Giant Magellan Telescope, in Chile, and the Thirty Meter Telescope in Hawaii. In space the field will see the successor of the Hubble Space Telescope (HST), the James Webb Space Telescope (JWST), launching in 2021. Finally, {\bf optical polarimetry} has become an important counterpart to VHE observations, as polarimetric observations offer an ideal way of isolating the synchrotron/non-thermal component from the total optical emission~\cite{Barres2014, Jermak2016}.

To conclude, the {\bf hard X-ray/soft gamma-ray domain} (0.1 - 10 MeV), and the {\bf high-energy gamma-ray GeV domain} represent important complements to CTA, providing immediate extrapolations of the VHE spectrum to lower energies. From the first group, INTEGRAL is the best example, which will hopefully continue to operate beyond this decade. In the MeV-GeV domain, Fermi-LAT is a primary space-based partner to CTA, on which CTA will greatly rely for the study of transients. The newest missions DAMPE and HERD (exp. 2020) come as useful partners above 100 GeV, and other proposed missions from EU and the US may also come online in the next decade. Finally, the ground-based particle HAWC and LHAASO constitute VHE all-sky complements to CTA operation, to which a southern version is highly expected.

\section{Multi-messenger and transients follow-up with CTA}

Transient astrophysical sources constitute the most demanding and critical sources in terms of MWL synergies for CTA. These objects are also potential sources for non-photonic signals, being at the central stage of the new multi-messenger astronomy. Due to their variable character, such objects have always been challenging to study in the VHE domain, but this picture is set to change dramatically, thanks to the unprecedented sensitivity of CTA at the lowest variability timescales~\cite{Funk2013}. An excellent example is the recent detection by MAGIC of the first gamma-ray burst (GRB) at sub-TeV energies~\cite{GRB}. The relatively large field of view (FoV) of CTA is another factor which will be crucial for transients and in allowing for the discovery of serendipitous sources by CTA. A real-time analysis (RTA) service is planned for CTA to recognise new transients and issue automatic alerts for follow ups by other telescopes~\cite{Fioretti2015}.

CTA proposes to follow up on observations of transient targets, among which are AGN, GRBs, Galactic transients, as well as the most interesting multi-messenger neutrinos and gravitational wave (GW) events. The time toll within CTA for such observations is expected to be quite large with aggregate observation times of up to 390 h/yr/site during the early phase operations of CTA~\cite{CTA2019}. But the requests from external facilities which will provide complementary MWL follow-ups is equally demanding, and will require strong cooperation for simultaneous or coordinated multi-instrumental campaigns in both hemispheres, and in space. In this context, follow-up campaigns of high-energy neutrino transients or GW events are top priority. Neutrino observatories such as IceCube and KM3Net (in construction) will provide regular alerts of high-energy neutrino events~\cite{Aartsen2013}, which are clear indicators of hadronic cosmic-ray production~\cite{Ahlers2015}. CTA follow-up of appropriately selected alerts (estimated in 2-3 per year) can give important insights on the origin of Galactic and extragalactic cosmic-rays~\cite{IceCube2018}. Equal priority will be given to the search of VHE counterparts to GW signal events provided by observatories such as VIRGO and Advanced LIGO~\cite{Abbott2016}. While no counterpart has yet been detected, the MAGIC GRB detection and the Fermi observations of a NS-NS merger~\cite{Abbott2017} provide indications that such goal is achievable by CTA.


\begin{thebibliography}{99}
\bibitem{Aartsen2013} Aartsen M. et al. 2013. Evidence for high-energy extraterrestrial neutrinos at the IceCube detector {\it Science}, {\bf 342},1242856.
\bibitem{Abbott2016} Abbott B.P., et al. 2016. Observations of gravitational waves from a binary black hole merger {\it Phys. Rev. Lett.}, {\bf 116},061102.
\bibitem{Abbott2017} Abbott B.P., et al. 2017. Multi-messenger observations of binary neutron star merger {\it ApJ Lett.}, {\bf 848}, 2.
\bibitem{Aharonian2006} Aharonian F., Akhperjanian, A.G., Bazer-Bachi, A.R. et al. 2006. Discovery of very high-energy gamma-rays from the Galactic Centre ridge. {\it Nature}, {\bf 439}, 695.
\bibitem{Aharonian2009} Aharonian F., Akhperjanian, A.G., Anton, G. et al. 2009. Spectrum and variability of the Galactic Center VHE gamma-ray source H.E.S.S. J1745-290. {\it A\&A}, {\bf 503}, 817.
\bibitem{Ahlers2015} Ahlers M. \& Halzen F. 2015. High-energy cosmic neutrino puzzle. {\it Rep. Prog. Phys.}, {\bf 78}, 126901.
\bibitem{Arsioli2018} Arsioli B., Barres de Almeida U., Prandini E. et al. 2018. Extreme- and high-synchrotron-peaked blazars at the limit of Fermi-LAT detectability. {\it MNRAS}, {\bf 480}, 2165.
\bibitem{Barres2014} Barres de Almeida U. et al. 2014. Polarimetric tomography of blazar jets {\it MNRAS}, {\bf 441}, 2885.
\bibitem{Bell2004} Bell A.R. 2004. Turbulent amplification of magnetic field and diffusive shock acceleration of cosmic rays  {\it MNRAS}, {\bf 353}, 550.
\bibitem{CTA2019} The CTA Consortium, 2019. Science with the Cherenkov Telescope Array (World Scientific), ISBN 978-981-3270-08-4, 338 pp, E-print: arXiv:1709.07997.
\bibitem{Drury1983} Drury L.O. 1983. An introduction to the theory of diffusive shock acceleration of energetic particles in tenuous plasmas  {\it Rep. Prog. Phys.}, {\bf 46}, 973.
\bibitem{Fioretti2015} Fioretti V. et al. 2015. RTA sensitivity evaluation of the Cherenkov Telescope Array  {\it Proc. 34th ICRC}.
\bibitem{Funk2013} Funk S. \& Hinton J, 2013. Comparison of Fermi-LAT and CTA between 10-100 GeV  {\it A.Ph.}, {\bf 43}, 348.
\bibitem{HESS2016} H.E.S.S. Collaboration, Abramowski A., Aharonian F. et al. 2016. Acceleration of petaelectronvolt protons in the Galactic Centre. {\it Nature}, {\bf 531}, 476.
\bibitem{Hillas2005} Hillas A.M. 2005. Can diffusive shock acceleration in supernova remnants account for high energy galactic cosmic rays? {\it Journal Phys. G}, {\bf 31}, 95.
\bibitem{IceCube2018} The IceCube Collaboration et al. 2018. Multi-messenger observations of a flaring blazar coincident with high-energy neutrino IceCube-170922A {\it Science}, {\bf 361}, 6398.
\bibitem{Jermak2016} Jermak H. et al. 2016. RINGO2 and DIPOL optical polarisation blazar catalogue {\it MNRAS}, {\bf 462}, 4267.
\bibitem{GRB} Mirzoyan R. \& The MAGIC Collaboration 2019. First-time detection of a GRB at sub-TeV energies; MAGIC detects the GRB 190114C {\it ATel} 12390.
\bibitem{Sol2013} Sol H. et al. 2013. Active Galactic Nuclei under the scrutiny of CTA {\it Astropart. Phys.}, {\bf 43}, 215.

\end{thebibliography}
\end{document}